\documentclass{elsart}
\usepackage{epsfig}
\usepackage{enumerate}

\begin{document}
\begin{frontmatter}

\title{The influence of contrarians in the dynamics of opinion formation}

\author{Joao Paulo Gambaro,}
\thanks{jpgambaro@if.uff.br}
\author{Nuno Crokidakis}
\thanks{nuno@if.uff.br}

\address{
Instituto de F\'{\i}sica, Universidade Federal Fluminense, Niter\'oi - Rio de Janeiro, Brazil}

\maketitle

\begin{abstract}
\noindent
In this work we consider the presence of contrarian agents in discrete 3-state kinetic exchange opinion models. The contrarians are individuals that adopt the choice opposite to the prevailing choice of their contacts, whatever this choice is. We consider binary as well as three-agent interactions, with stochastic parameters, in a fully-connected population. Our numerical results suggest that the presence of contrarians destroys the absorbing state of the original model, changing the transition to the para-ferromagnetic type. In this case, the consequence for the society is that the three opinions coexist in the population, in both phases (ordered and disordered). Furthermore, the order-disorder transition is suppressed for a sufficient large fraction of contrarians. In some cases the transition is discontinuous, and it changes to continuous before it is suppressed. Some of our results are complemented by analytical calculations based on the master equation.

\end{abstract}
\end{frontmatter}

Keywords: Opinion Dynamics, Collective Phenomenon, Phase Transition, Agent-based model, Computer Simulation

\section{Introduction}

\qquad Dynamics of social systems have been studied using the statistical physics approach in the last thirty years. Indeed, different kinds of social phenomena emerge from the mutual influence of a large number of individuals, making possible a simple statistical physics modeling \cite{galam_book,sen_book,loreto_rmp}. From the theoretical point of view, social dynamics' models are interesting to physicists because they present scaling, universality, long-range correlations and order-disorder transitions, among other typical features of statistical physics systems \cite{loreto_rmp}.
 
The Galam's majority-rule model is one of the most famous models for study the process of opinion formation and evolution \cite{galam,galam1}. In the standard formulation of such model, groups of 3 agents are randomly chosen in a population, and after the interaction all agents follow the local majority. The model leads to consensus states where all agents in the society share one of the two possible opinions, depending on the initial concentrations of such opinions.   

One interesting modification of the majority-rule model was to include a class of special agents called contrarians \cite{galam2,galam_contrarians}. The contrarians are individuals that adopt the choice opposite to the prevailing choice of his contacts, whatever this choice is. As a consequence of the presence of such special individuals, it was verified that the consensus states do not occur anymore, with the occurrence of a mixed phase with the coexistence of both opinions, where one of the opinions dominates, i.e., we have an ordered phase. Galam also verified the occurrence of a phase transition at a critical density of contrarians $a_{c}=1/6$. In this phase, he discussed that \textit{``agents keep shifting states but no global symmetry breaking, i.e., the appearance of a majority, takes place''} \cite{galam_contrarians}. Those contrarian individuals were also considered in a series of papers concerning models of opinion dynamics \cite{pre_pmco,allan_physA,lalama1,lalama2,jorge,tanabe,masuda,biswas,bagnoli,jarman,guo}.

In this work we introduce the Galam's contrarians in kinetic exchange opinion models. Our target was to analyze the critical behavior of the model. We follow a similar approach of Galam, i.e., we consider a density $a$ of contrarian attitudes in the population. One the other hand, in the case of conformist (non-contrarian) agents, the interactions occur through kinetic exchanges with stochastic parameters. We considered both pairwise and three-agent interactions. 

The paper is organized as follows. In Section 2 we present the first formulation of the model, considering contrarians and pairwise interactions. In Section 3 we discuss the results for the second formulation of the model, considering contrarians and interactions in groups formed by three agents. The final remarks are  presented in Section 4.


\section{Model I: Pairwise interactions and contrarians}

\qquad Our model is based on the kinetic exchange opinion model of references \cite{pre_biswas,lccc}. We consider a fully-connected population of size $N$, where the agents interact by pairs. Each individual $i$ ($i=1,2,...,N$) carries one of three possible opinions, represented by $o_{i}=+1, -1$ or $0$. This scenario  mimics any polarized  public debate, for example an electoral process with two different candidates A and B, where each agent (or elector) votes for the candidate A (opinion $+1$), for the candidate B (opinion $-1$) or remains undecided/neutral (opinion $0$) \cite{allan_physA,wu,balakin}. In addition to a kinetic exchange rule of interaction, we consider a fraction $a$ of contrarians in the population, similarly to the approach of Galam in the majority-rule model \cite{galam_contrarians}. Each interaction occurs between two given agents $i$ and $j$, such that $j$ will influence $i$. The following rules govern the dynamics at a given time step $t$:

\begin{enumerate}

\item A pair of agents $(i,j)$ is randomly chosen;

\item If $i$ is a non-contrarian (conformist) agent, his/her opinion in the next time step $t+1$ will be updated according to
\begin{equation}\label{eq1}
o_{i}(t+1) = {\rm sgn}[\lambda\,o_{i}(t)+\lambda\,\epsilon\,o_{j}(t)],
\end{equation}
where the sign function is defined such that ${\rm sgn}(0)=0$, and the stochastic variables $\lambda$ and $\epsilon$ will be defined in the following.  

\item On the other hand, if $i$ is a contrarian, his/her opinion in the next time step $t+1$ will be updated according to

\begin{itemize}  
  
\item If $o_{j}(t)=1$, then  $o_{i}(t+1)=-1$

\item If $o_{j}(t)=-1$, then $o_{i}(t+1)=1$

\item If $o_{j}(t)=0$, then $o_{i}(t+1)=1$ or $-1$ with equal probability ($1/2$)

\end{itemize}    
  
\end{enumerate}

In other words, with probability $a$ the individual $i$ will act as a contrarian agent. In this case he takes the opposite opinion of his neighbor $j$. If the neighbor $j$ is neutral, i.e., if $o_{j}=0$, agent $i$ takes one of the extreme opinions $\pm 1$ with equal probability. Otherwise, with probability $1-a$ agent $i$ will act as a conformist agent. Thus, agent $j$ will influence agent $i$ through the kinetic exchange rule of Eq. (\ref{eq1}), where the stochastic variables $\lambda$ and $\epsilon$ are given by the discrete probability distributions $F$ and $G$, given respectively by
\begin{eqnarray} \label{eq2}
F(\lambda) & = & p\,\delta(\lambda-1) + (1-p)\,\delta(\lambda) ~, \\ \label{eq3}
G(\epsilon) & = & \frac{1}{2}\,[\delta(\epsilon-1) + \,\delta(\epsilon)] ~.
\end{eqnarray}

The interpretation of Eqs. (\ref{eq1}-\ref{eq3}) is simple. If agent $i$ is a conformist individual, after the interaction he retains a fraction of his own opinion (which depends on the agent's conviction $\lambda$, that is the same for all agents, for simplicity) and is stochastically influenced by agent $j$ (which depends on the variable $\epsilon$). To facilitate the analytical treatment, we consider that the variables $\lambda$ and $\epsilon$ are annealed random variables, i.e., at each interaction we choose their values following the distributions (\ref{eq2}) and (\ref{eq3}). In addition, every time we choose an agent $i$, their state, contrarian or conformist, is defined with probabilities $a$ and $1-a$, respectively, as was done in the Galam's majority-rule model \cite{galam_contrarians}. In this sense, $a$ is also an annealed variable. \footnote{The quenched version consider that a fraction $a$ of the population is randomly selected as contrarians at the beginning of the simulation, and they keep this character throughout the dynamics.}

To start the analysis of the model, we consider the order parameter $O$, that gives us the average opinion of the agents, and is defined by $O=\frac{1}{N}\,|\sum_{i=1}^{N}\,o_{i}|$. One can enumerate all the processes that contribute to the increase or decrease of the order parameter. Thus, following \cite{pre_biswas}, the master equation for $O$ can be given by
\begin{eqnarray}\nonumber
\frac{d}{dt}\,O & = & [2a+(1-a)(1-p)]f_{-1}^{2}+\frac{1}{2}af_{0}^{2}+\frac{1}{2}(1-a)pf_{0}f_{1} \\ \nonumber
&   &\mbox{}+[2a+(1-a)(1-p)]f_{0}f_{-1}-[2a+(1-a)(1-p)]f_{1}^{2} \\ \label{eq4}
&   &\mbox{}-\frac{1}{2}af_{0}^{2}-\frac{1}{2}(1-a)pf_{0}f_{-1}-[2a+(1-a)(1-p)]f_{0}f_{-1}
\end{eqnarray}
where $f_{1},f_{-1}$ and $f_{0}$ denote the fractions or densities of opinions $+1, -1$ and $0$, respectively. In the stationary state $dO/dt=0$. Using the normalization condition $f_{1} + f_{-1} + f_{0} = 1$, we obtain two solutions for Eq. (\ref{eq4}) in the stationary state, namely $2f_{1}+f_0=1$, which implies in $f_{1}=f_{-1}=(1-f_{0})/2$ (disordered paramagnetic state), or
\begin{equation}\label{eq5}
f_{0} =  \frac{2[2a + (1-a)(1-p)]}{(1-a)p} ~.
\end{equation}

In this case, Eq. (\ref{eq5}) is valid in the ferromagnetic ordered phase. We emphasize that if we take $a=0$ in Eq. (\ref{eq5}) one obtains $f_{0} = 2(1-p)/p$, which agrees with the result of Ref. \cite{pre_biswas}. In ref. \cite{pre_biswas}, there is a critical point $p_{c}=2/3$ that separates an absorbing phase (for $p\leq p_{c}$) from a ferromagnetic ordered phase (for $p> p_{c}$). Thus, the first identified effect of the presence of contrarians is change the nature of the phase for $p\leq p_{c}$. Indeed, we found a paramagnetic state solution $f_{1}=f_{-1}=(1-f_{0})/2$, that will be discussed in more details in the following. Notice that in our case the values of $p_{c}$ will depend on $a$, as we will see in the following.

To better study the effect of contrarians on the mentioned change of nature (absorbing to paramagnetic), one can consider the fluxes into/out the neutral state $o=0$. One can write a master equation for $f_{0}$ considering the above disordered solution $f_{1}=f_{-1}=(1-f_{0})/2$. In this case, we have
\begin{equation} \label{eq6_new}
\frac{d}{dt}\,f_{0}=\frac{3}{4}p(a-1)f_{0}^{2}+f_{0}+\left(\frac{3}{4}p-1\right)(1-a) ~.  
\end{equation}
Thus, in the stationary state we have $df_{0}/dt=0$, and Eq. (\ref{eq6_new}) gives us
\begin{equation} \label{eq7_new}
f_{0}=\frac{2-\sqrt{9a^{2}p^{2}-12a^{2}p-18ap^{2}+24ap+9p^{2}-12p+4}}{3p(1-a)} ~.
\end{equation}

In this case, Eq. (\ref{eq7_new}) is valid in the paramagnetic disordered phase. One can see from Eq. (\ref{eq7_new}) that we have $f_{0}(a=0)=1$, which recover the absorbing state solution of Ref. \cite{pre_biswas}, since the normalization condition leads to $f_{1}=f_{-1}=0$. However, for $a\neq 0$ we have $f_{0}<1$, and $f_{1}=f_{-1}=(1-f_{0})/2\neq 0$, i.e., a paramagnetic disordered state, confirming the above discussion about the impact of contrarians. Thus, the contrarian behavior leads to the coexistence of the three opinions in the region $p<p_{c}(a)$, but in that region there is no dominating opinion ($f_{1}=f_{-1}$).

Let us focus our attention in the ordered phase, i.e., in the region $p>p_{c}(a)$. In order to obtain an analytical expression for the order parameter $O$, one can consider the fluxes into/out the state $o=+1$. The master equation for $f_{1}$ is then
\begin{eqnarray} \nonumber
\frac{d}{dt}\,f_{1} & = & \frac{3}{2}af_{-1}f_{0}+af_{-1}^{2}+\frac{1}{2}(1-a)pf_{0}f_{1}+\frac{1}{2}af_{0}^{2}-\frac{3}{2}af_{-1}f_{0}+af_{-1}^{2}  \\  \label{eq6}
&   &\mbox{}+\frac{1}{2}(1-a)pf_{0}f_{1}+\frac{1}{2}af_{0}^{2} ~.
\end{eqnarray}

Considering the normalization condition $f_{1} + f_{-1} + f_{0} = 1$ and the expression for $f_{0}$ valid in the ferromagnetic phase, Eq. (\ref{eq5}), we obtain for $f_{1}$ in the stationary state (where $df_{1}/dt=0$)
\begin{eqnarray}  \label{eq7}
f_{1} = \frac{3p(1-a)-2(a+1)\pm \sqrt{\Delta}}{2(1-a)p} ~,
\end{eqnarray}
\noindent
where 
\begin{equation}  \label{eq8}
\Delta=9a^{2}p^{2}+28a^{2}p-18ap^{2}+12a^{2}-16ap+9p^{2}+16a-12p+4 ~.
\end{equation}

One can see that Eq. (\ref{eq7}) predicts two solutions (see the $\pm$ signals), i.e., one has two curves as functions of $p$ for each value of $a$. When $f_{1}$ assumes one of these values, consequently $f_{-1}$ takes the other one \cite{allan_physA}. Indeed, one can calculate separately the stationary fraction $f_{-1}$. From the normalization condition $f_{1} + f_{-1} + f_{0} = 1$, together with the expression for $f_{0}$ valid in the ferromagnetic phase, Eq. (\ref{eq5}), we have
\begin{eqnarray}\label{eq9}
f_{-1} = \frac{3p(1-a)-2(a+1)\mp \sqrt{\Delta}}{2(1-a)p} ~.
\end{eqnarray}

Thus, as Eq. (\ref{eq7}), Eq. (\ref{eq9}) indicates that when $f_{-1}$ assumes one of the values ($+$ or $-$ signal), consequently $f_{1}$ takes the other one. One can see from Eqs. (\ref{eq5}), (\ref{eq7}) and (\ref{eq9}) that the 3 states ($+1$, $-1$ and $0$) also coexist in the population in the ordered phase. This is another consequence of the introduction of contrarians in the population. Indeed, in the absence of contrarians ($a=0$) one of the extreme opinions disappears of the population in the ordered phase, and the surviving opinion coexists with neutral/undecided agents \footnote{This result can also be seen putting $a=0$ in Eqs. (\ref{eq7}) and (\ref{eq9}).} \cite{pre_biswas}. Thus, our formulation of the model is more realistic in the sense that distinct opinions can coexist in the population, as occurs usually in referendums and elections \cite{galam_book,sen_book}, and there is a dominating opinion, as we will see in the following.

Finally, the order parameter can be found from the relation $O=|f_{1}-f_{-1}|$. Considering Eqs. (\ref{eq7})-(\ref{eq9}), one obtains
\begin{eqnarray} \label{eq10}
O(a,\,p)={\frac{\sqrt{\Delta}}{p\left(1-a\right)}} ~.
\end{eqnarray}
where $\Delta$ is given by Eq. (\ref{eq8}). From Eq. (\ref{eq8}) we have $\Delta(a=0)=9p^{2}-12p+4=(3p-2)^{2}$, and Eq. (\ref{eq10}) gives us $O(0,p)=3(p-2/3)/p$, recovering the result of ref. \cite{pre_biswas}. The critical points $p_{c}(a)$ can be found from Eq. (\ref{eq10}) taking $O(a,\,p)=0$, which gives us
\begin{equation} \label{eq11}
p_{c}(a)=\frac{2}{9}\frac{\sqrt{2a(11a+3)}+7a+3}{(1-a)} ~.
\end{equation}

Notice that the Eq. (\ref{eq11}) recovers the result $p_{c}(a=0)=2/3$ of ref. \cite{pre_biswas}. Looking for the above results, one can see that they predict a paramagnetic-ferromagnetic transition, as discussed previously. Indeed, the result $O(0,p)=3(p-2/3)/p$ can be compared with the usual relation $O\sim (p-p_{c})^{\beta}$, and we obtain $\beta=1$, a typical exponent of an active-absorbing transition \cite{dickman}, as discussed in \cite{pre_biswas}. However, in the presence of contrarians, our result for the order parameter $O(a,p)$, Eq. (\ref{eq10}), gives us $O\sim (p-p_{c}(a))^{1/2}$, i.e., we have $\beta=1/2$. In other words, a typical mean-field Ising exponent, that describes the behavior of the order parameter near a para-ferromagnetic phase transition.

\begin{figure}[t]
\begin{center}
\vspace{0.2cm}
\includegraphics[width=0.7\textwidth,angle=0]{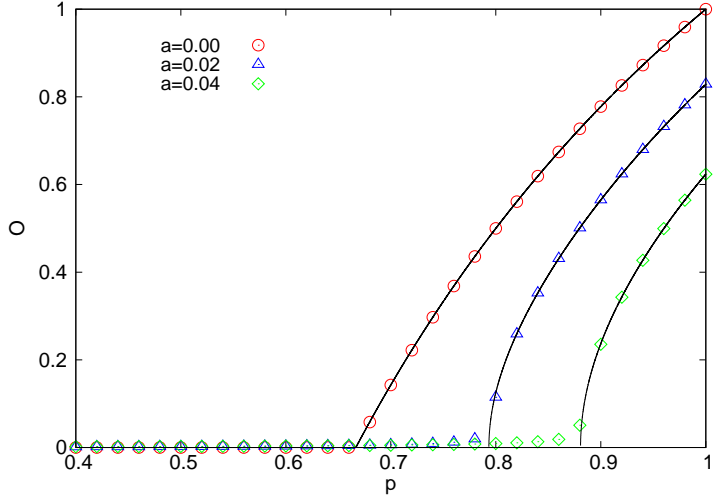}
\end{center}
\caption{(Color online) Order parameter $O$ as a function of the probability $p$ for typical values of the contrarian probability $a$ for the model with pairwise interaction. One can see phase transitions at distinct values of $p_{c}$ that depend of $a$. The symbols are numerical simulations for population size $N=2\times 10^{4}$, averaged over $100$ independent simulations. The full lines are the analytical predictions, Eq. (\ref{eq10}).}
\label{fig1}
\end{figure}

To complement our results, we performed numerical simulations for a population size $N=2\times 10^{4}$. We started all simulations with a fully-disordered population ($1/3$ of each opinion). One time step in the model is defined as the application of the above-mentioned rules $N$ times. We considered asynchronous (sequential) updates of agents' opinions. We computed the order parameter $O$ by the relation $O=\frac{1}{N}\,|\sum_{i=1}^{N}\,o_{i}|$. In Fig. \ref{fig1} we exhibit the numerical results for $O$ versus $p$ and typical values of $a$, together with the analytical result given by Eq. (\ref{eq10}). One can observe transitions at different points $p_{c}$ that depend on $a$, with the usual finite-size effects for $a\neq 0$, which is expected in order-disorder transitions. For the case $a=0$, as we have an absorbing state for $p<p_{c}$ we do not expect such finite-size effects. It is exactly which we see in Fig. \ref{fig1} for $a=0$.

As one can also see in Fig. \ref{fig1}, the population does not reaches consensus for $a\neq 0$, even for $p=1$. The order parameter presents a maximum value for $p=1$, but this value decreases for increasing values of $a$, suggesting that the transition may be suppressed for sufficient high values of the probability $a$. The critical value $a_{c}$ above which there is no more ordering in the population, i.e., where the model does not present the phase transition anymore, may be found taking the limiting case $p_{c}=1$ in Eq. (\ref{eq11}). In this case, one obtains
\begin{equation}\label{eq12}
a_{c}=\frac{9-4\sqrt{2}}{49}\approx 0.068 ~.
\end{equation}

\begin{figure}[t]
\begin{center}
\vspace{0.5cm}
\includegraphics[width=0.46\textwidth,angle=0]{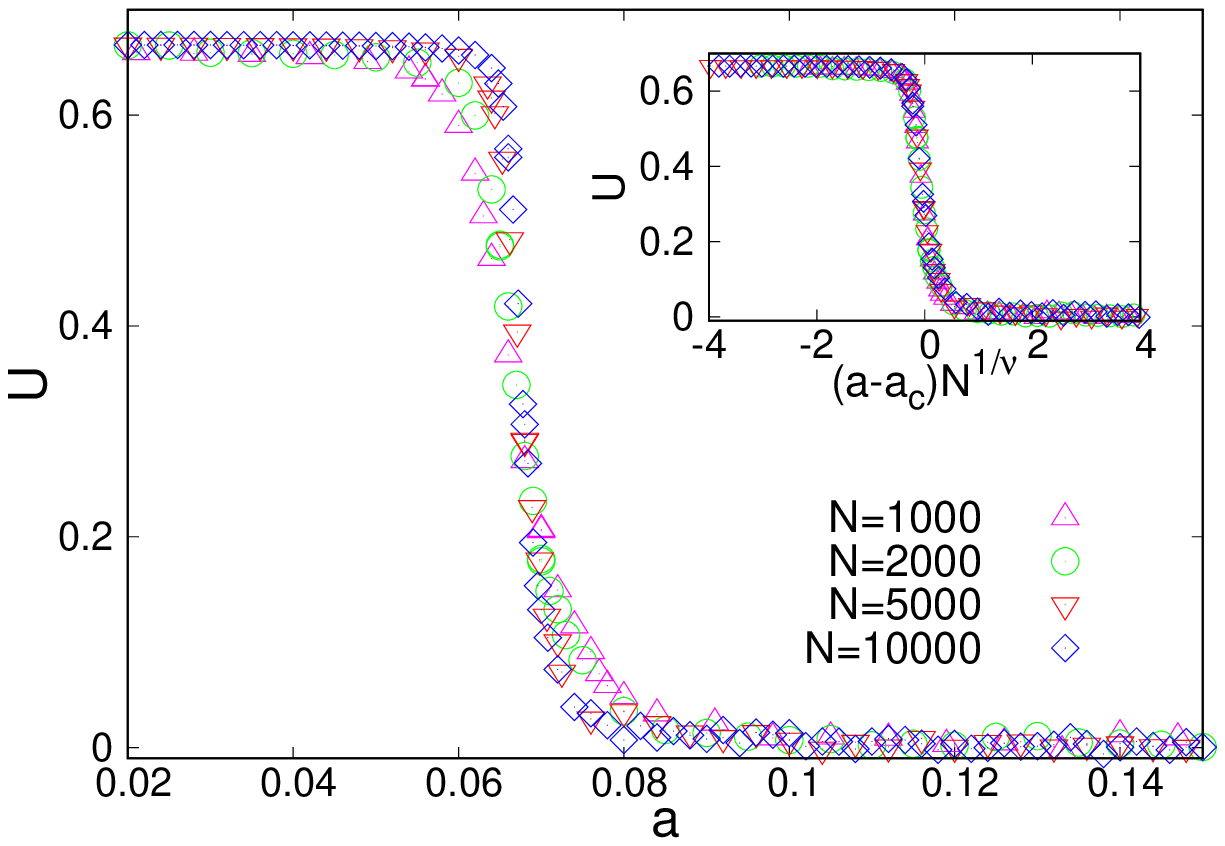}
\hspace{0.5cm}
\includegraphics[width=0.46\textwidth,angle=0]{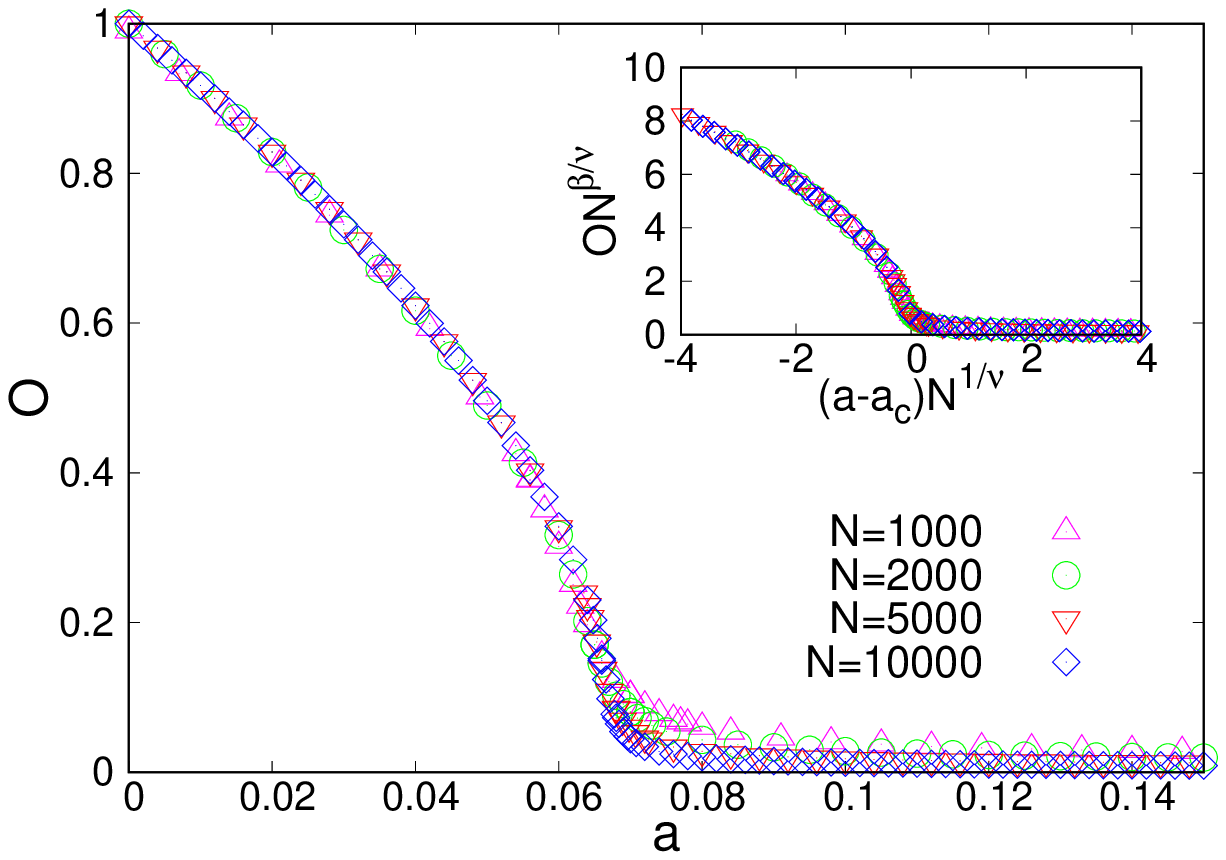}
\\
\includegraphics[width=0.46\textwidth,angle=0]{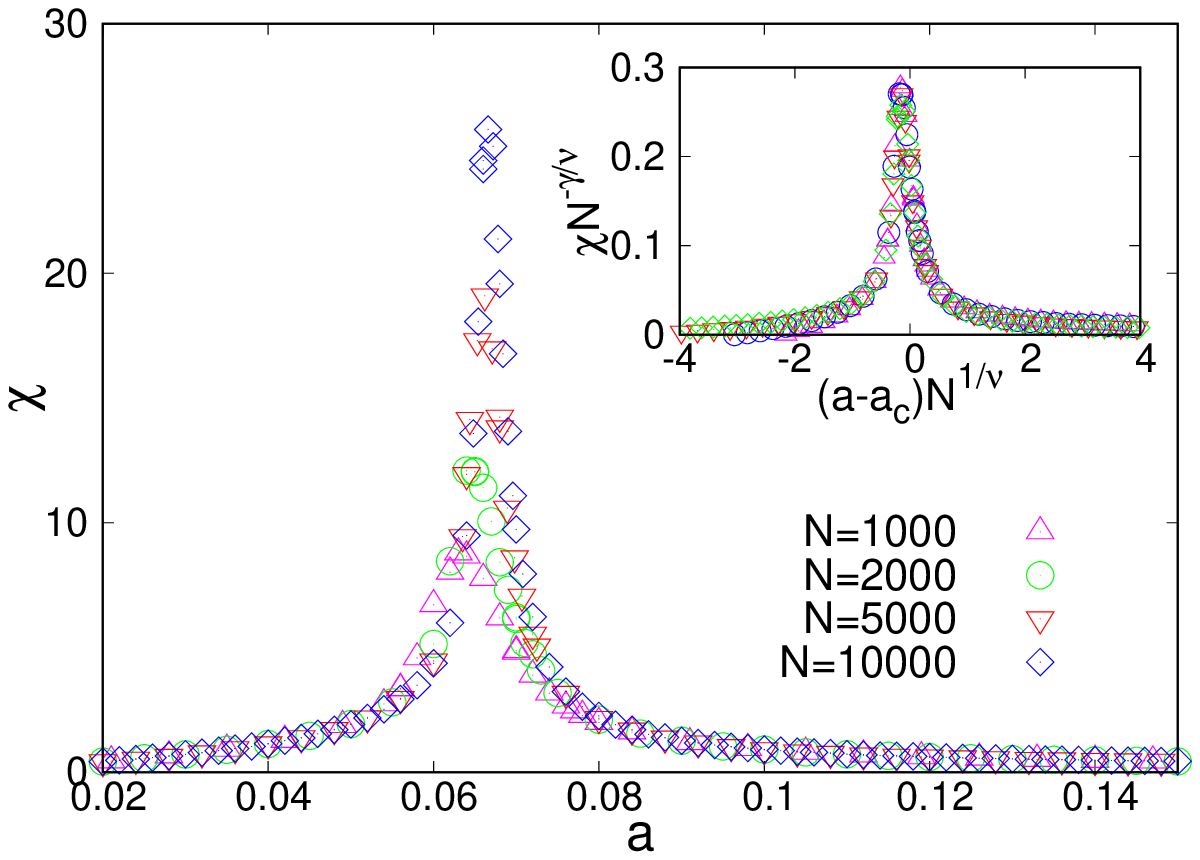}
\end{center}
\caption{(Color online) Binder cumulant $U$ (upper, left panel), order parameter $O$ (upper, right panel) and susceptibility $\chi$ (lower) as functions of $a$ for the limiting case $p=1$ and distinct population sizes $N$ for the model with pairwise interaction. The crossing of the Binder cumulant curves occurs for $a_{c}\approx 0.068$, which agrees with the analytical result, Eq. (\ref{eq12}). In addition, the insets present the data collapses. The estimated critical exponents are $\beta\approx 0.5$, $\gamma \approx 1$ and $\nu\approx 2$.}
\label{fig2}
\end{figure}

Thus, if we have at least about $7\%$ of contrarian attitudes in the population, the long-time behavior of the public debate is given by a disordered state, i.e., there is no majority or dominating opinion in the population, with both state densities equal (zero order parameter). On the other hand, if the fraction of contrarians is less than $7\%$, one of the opinions ($+1$ or $-1$) are dominant in the population. One can corroborates this result performing numerical simulations of the model for $p=1$. In Fig. \ref{fig2} we exhibit results for the Binder cumulant $U$ (upper figure, left side), the order parameter $O$ (upper figure, right side) and the ``susceptibility'' $\chi$ (lower figure) as functions of $a$ for $p=1$ and distinct population sizes $N$. The Binder cumulant is defined as $U=1-\langle O^{4}\rangle/3\langle O^{2}\rangle^{2}$, whereas the susceptibility is given by $\chi=N\,(\langle O^{2}\rangle-\langle O\rangle^{2})$. One can see that the order parameter goes to zero (with finite-size effects) in the range $0.06 < a < 0.09$. In addition, one can see a crossing of the Binder cumulant curves at $a\approx 0.068$, which is in agreement with the analytical result of Eq. (\ref{eq12}). Furthermore, we also exhibit in the insets of Fig. \ref{fig2} the corresponding scaling plots, obtained from standard finite-size scaling relations,
\begin{eqnarray} \label{eq13}
O(N) & \sim & N^{-\beta/\nu} ~, \\  \label{eq14}
\chi(N) & \sim & N^{\gamma/\nu} ~, \\   \label{eq15}
a-a_{c} & \sim & N^{-1/\nu} ~.
\end{eqnarray}
The best data collapse was obtained for $a_{c}\approx 0.068$, $\beta\approx 0.5$, $\gamma \approx 1$ and $\nu\approx 2$. Thus, typical mean-field Ising exponents, which is expected due to the mean-field character of the model.


\section{Model II: Three-agent interactions and contrarians}

\qquad The second formulation of the model considers interactions in a group of size $3$. Again, in addition to a kinetic exchange rule of interaction, we consider a fraction $a$ of contrarians in the population. Each interaction occurs among three given agents $i$, $j$ and $k$, such that the pair $(j,k)$ will influence $i$. The following rules govern the dynamics at a given time step $t$:

\begin{enumerate}

\item Three agents $(i,j,k)$ are randomly chosen;

\item If $i$ is a non-contrarian (conformist) agent, his/her opinion in the next time step $t+1$ will be updated according to
\begin{equation}\label{eq16}
o_{i}(t+1)={\rm sgn}[\lambda o_{i}(t)+\epsilon\lambda o_{jk}(t)] ~.
\end{equation}
where \[
o_{jk}(t)=\left\{ \begin{array}{c}
o_{j}(t)\,\,\,if\,\,o_{j}(t)=o_{k}(t)\\
0\,\,\,\,\,\,\,\,\,\,\,\,\,\,\,\,\,\,\,\, otherwise
\end{array}\right.
\]

and the stochastic variables $\lambda$ and $\epsilon$ are defined as in the previous subsection, see Eqs. (\ref{eq2}) and (\ref{eq3}).  

\item On the other hand, if $i$ is a contrarian, his/her opinion in the next time step $t+1$ will be updated depending on the opinions $o_{j}$ and $o_{k}$. We have two distinct cases:

\begin{itemize}  
  
\item Case I: $o_{j}=o_{k}$ 

\begin{itemize}  
  
\item If $o_{j}(t)\neq 0$, then $o_{i}(t+1)=-o_{j}(t)$

\item If $o_{j}(t)=0$, then $o_{i}(t+1)=1$ or $-1$ with equal probability ($1/2$)

\end{itemize}

\item Case II: $o_{j}\neq o_{k}$

\begin{itemize}  
  
\item If $o_{j}(t)=1$ and $o_{k}(t)=-1$, then $o_{i}(t+1)=0$

\item If $o_{j}(t)=-1$ and $o_{k}(t)=1$, then $o_{i}(t+1)=0$

\item If $o_{j}(t)=1$ and $o_{k}(t)=0$, then $o_{i}(t+1)=-1$

\item If $o_{j}(t)=0$ and $o_{k}(t)=1$, then $o_{i}(t+1)=-1$

\item If $o_{j}(t)=-1$ and $o_{k}(t)=0$, then $o_{i}(t+1)=1$

\item If $o_{j}(t)=0$ and $o_{k}(t)=-1$, then $o_{i}(t+1)=1$

\end{itemize}

\end{itemize}    
  
\end{enumerate}

Summarizing, the contrarian $i$ try to take the opposite opinion of the pair $(j,k)$, similarly to the dynamics of the majority-rule model in the presence of contrarians \cite{galam_contrarians}. As discussed in \cite{pre_biswas}, the master equation for the order parameter in the case of three-agent interaction presents terms of third order regarding the fractions of opinions, like $f_{1}^{2}f_{-1}$ or $f_{1}^{3}$. In ref. \cite{pre_biswas}, the author observes that one of the extreme opinions $1$ or $-1$ disappears of the population in the ordered phase, which allow him to take the above mentioned terms equal to zero, i.e., $f_{1}^{2}f_{-1}=0$, and others. In our case, we verified numerically that the the three opinions coexist in the population in the ordered phase. In this case, one cannot solve analytically the model in the presence of contrarians due to the mentioned terms involving products like $f_{1}^{2}f_{-1}$ (that are not zero in our case). In this case, we performed several simulations of the model, and the results are discussed in the following.

Following the discussion of the previous section, our first interest in the model is study the behavior of the order parameter as a function of $p$, for typical values of the contrarian probability $a$. Thus, we considered populations of size $N=2\times 10^{4}$. As the standard model \cite{pre_biswas} exhibits a discontinuous transition, we follow its analysis and performed simulations starting both from a fully-disordered ($1/3$ of each opinion) and a fully-ordered population (all agents sharing opinion $+1$ or $-1$), in order to analyze the occurrence of hysteresis behavior \cite{pre_2010}. One time step in the model is defined as the application of the above-mentioned rules $N$ times.

\begin{figure}[t]
\begin{center}
\vspace{6mm}
\includegraphics[width=0.46\textwidth,angle=0]{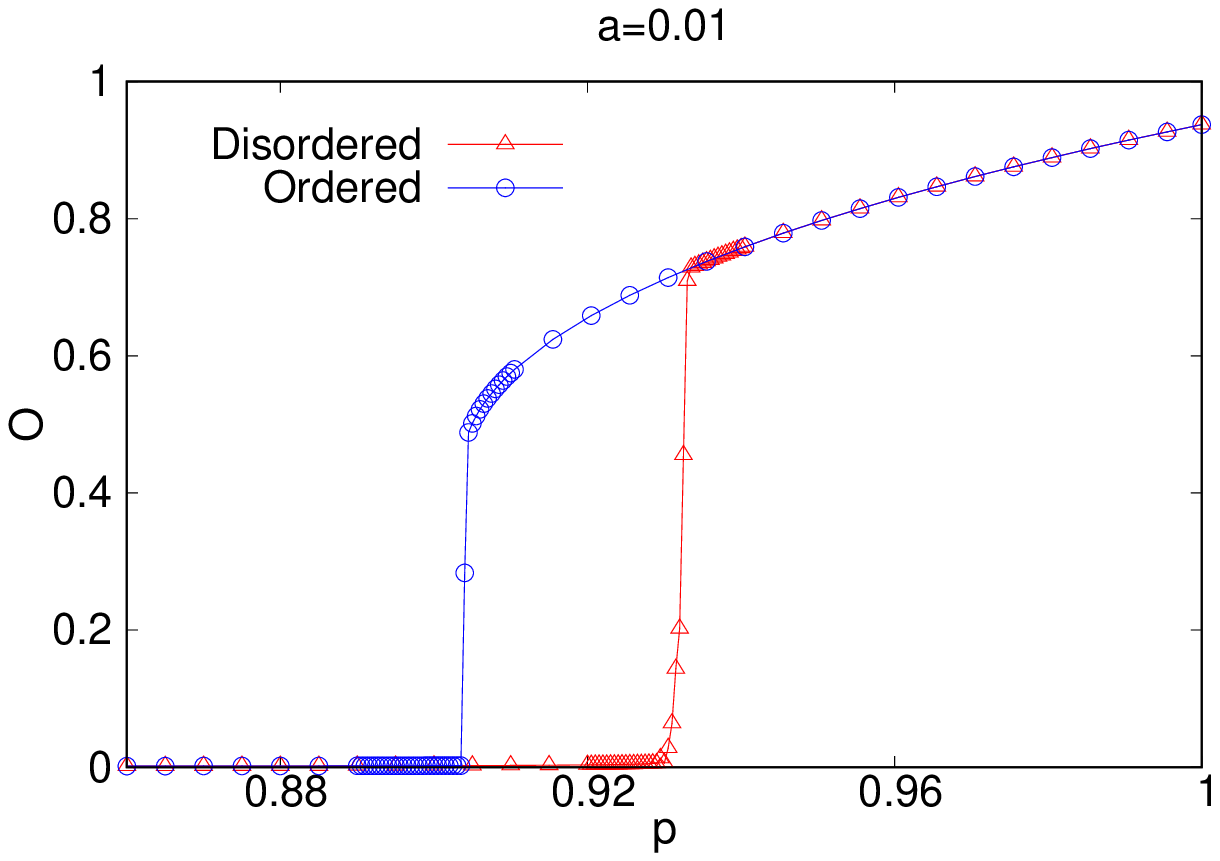}
\hspace{0.3cm}
\includegraphics[width=0.46\textwidth,angle=0]{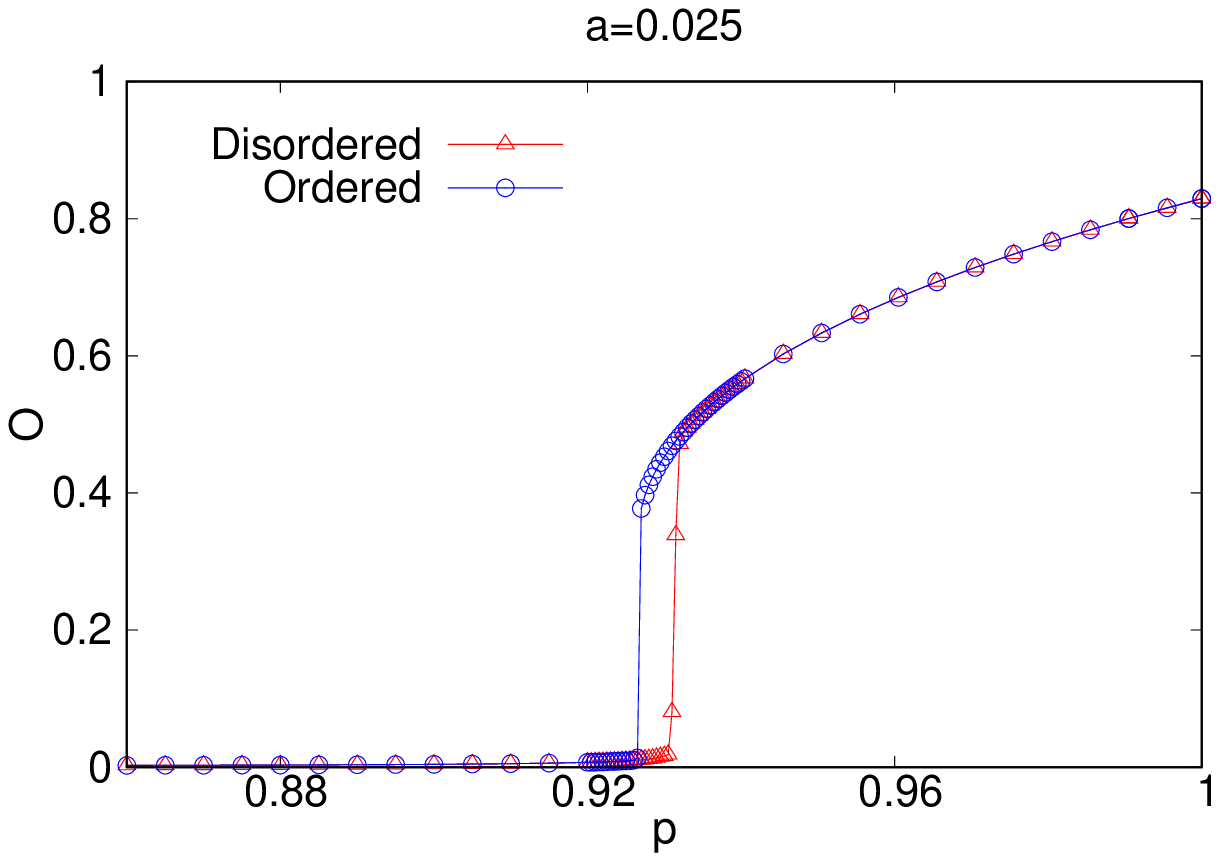}
\\
\vspace{0.5cm}
\includegraphics[width=0.46\textwidth,angle=0]{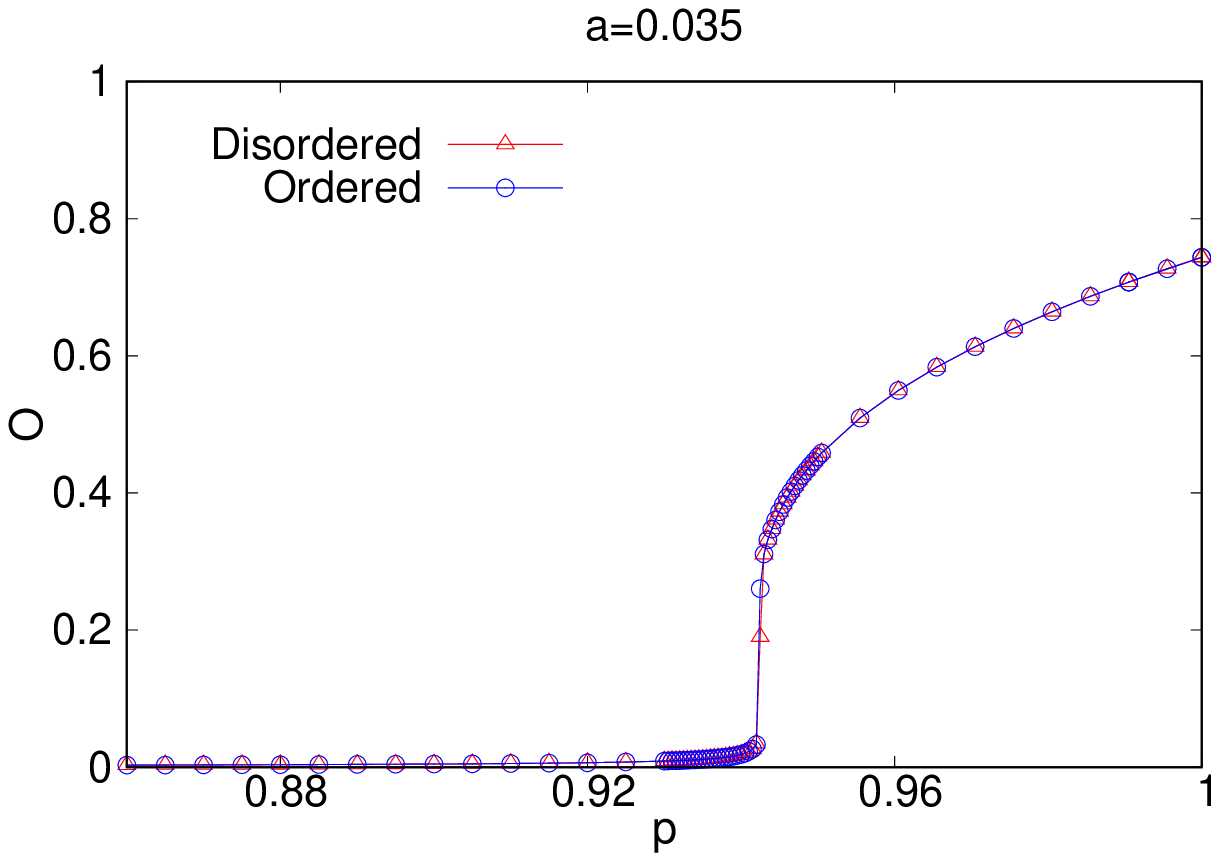}
\hspace{0.3cm}
\includegraphics[width=0.46\textwidth,angle=0]{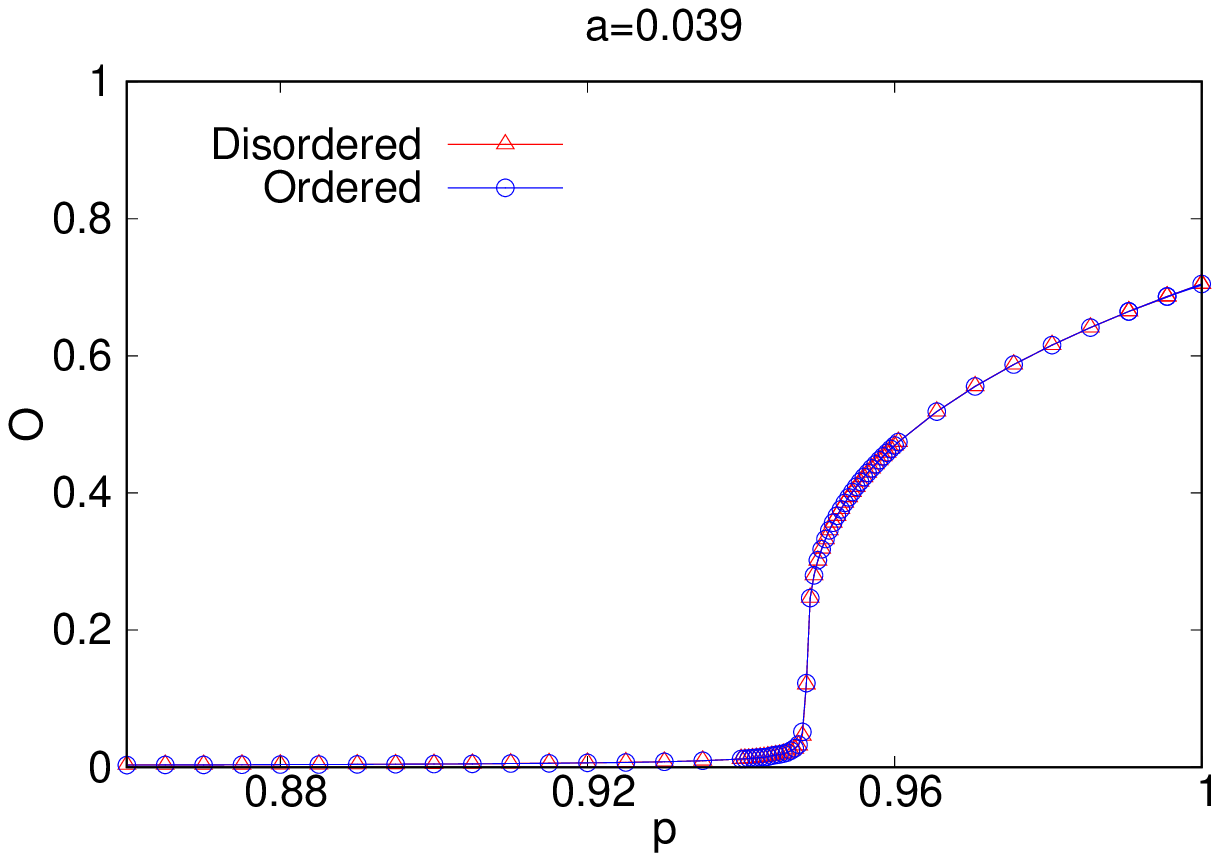}
\end{center}
\caption{(Color online) Order parameter as a function of the probability $p$ for typical values of the contrarian probability $a$ for the model with three-agent interaction. Each graphic exhibits data considering ordered and disordered initial conditions. One can see phase transitions at distinct values of $p_{c}$ that depend of $a$. The population size is $N=2\times 10^{4}$ and data are averaged over $100$ independent simulations.} 
\label{fig3}
\end{figure}

In Fig. \ref{fig3} we exhibit the numerical results considering the mentioned ordered and disordered initial conditions. For sufficient small values of $a$ like $a=0.01$ and $a=0.025$ one can see similar results in comparison with the model without contrarians, i.e., the hysteresis behavior typical of first-order phase transitions. However, the inclusion of contrarians shifts the critical points $p_{c}(a)$ to higher values in comparison with the original model \cite{pre_biswas}, and the area of the hysteresis decreases for raising $a$. In addition, we do not observe an absorbing phase for $p<p_{c}$ as in the original model. Instead we see a paramagnetic phase, as in the case with pairwise interactions analyzed in the previous section. In other words, we verified numerically the usual finite-size effects of paramagnetic phases in the region $p<p_{c}$, i.e., an order parameter slightly greater than zero. On the other hand, for higher values of $a$ like $a=0.035$ and $a=0.039$ the hysteresis does not occur anymore, and the transition becomes continuous. Thus, the model presents a tricritical point (TCP). Based on our numerical results, we monitored the occurrence of hysteresis behavior in order to estimate the location of such point. Our estimate is that $a_{TCP}$ is in the region $0.033<a_{TCP}<0.034$.

\begin{figure}[t]
\begin{center}
\vspace{0.5cm}
\includegraphics[width=0.46\textwidth,angle=0]{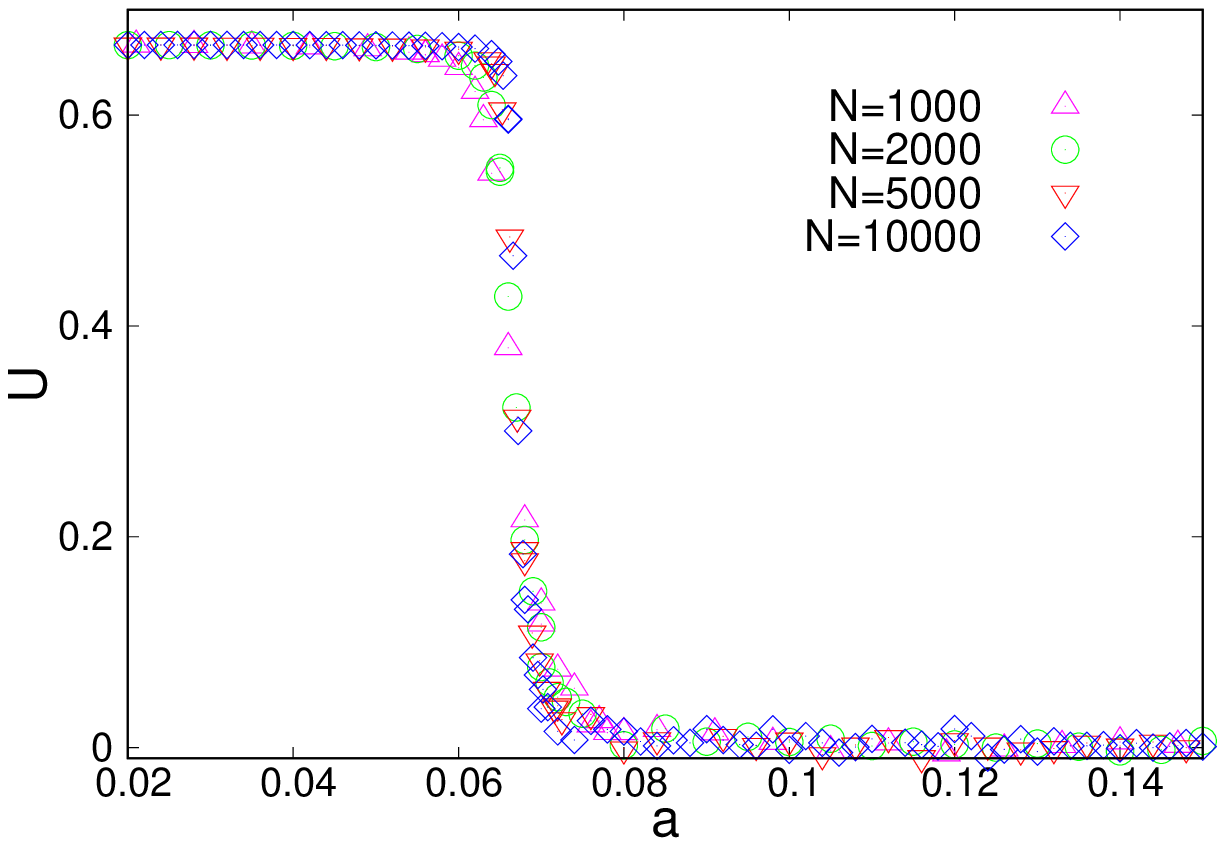}
\hspace{0.5cm}
\includegraphics[width=0.46\textwidth,angle=0]{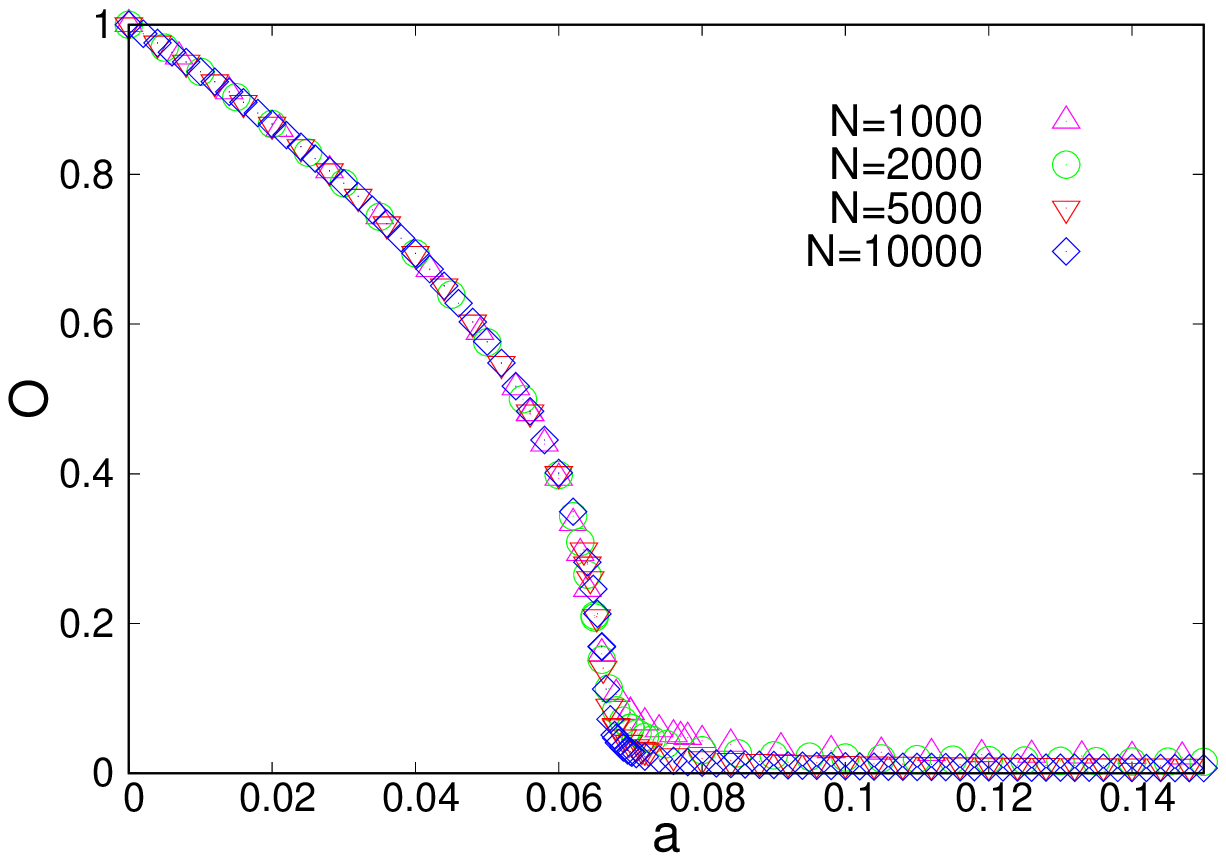}
\end{center}
\caption{(Color online) Binder cumulant $U$ (left panel) and order parameter $O$ (right panel) as functions of $a$ for the limiting case $p=1$ and distinct population sizes $N$ for the model with three-agent interaction. The crossing of the Binder cumulant curves occurs for $a_{c}\approx 0.068$.}
\label{fig4}
\end{figure}

As in the model with pairwise interaction, one can also see in Fig. \ref{fig3} that the values of the transition points $p_{c}(a)$ increases for raising $a$. Thus, in order to estimate the critical point $a_{c}$ above which there is no transition and the system is always disordered, we performed simulations for the limiting case $p=1$. In Fig. \ref{fig4} we show the results for the Binder cumulant (left side) and the order parameter (right side) as functions of $a$ for $p=1$ and distinct population sizes $N$. One can see  that the order parameter goes to zero (with finite-size effects) in the range $0.06 < a < 0.09$. In addition, one can see a crossing of the Binder cumulant curves at $a\approx 0.068$. It is the same value obtained in the case of pairwise interactions, suggesting that the critical value $a_{c}$ is robust against the type of interaction, at least considering both interactions among two or three agents. We also performed a finite-size scaling analysis (not shown), as in the previous case, and we found the same critical exponents, i.e., $\beta\approx 0.5$, $\gamma \approx 1$ and $\nu\approx 2$, supporting the mean-field-like exponents.

\section{Final Remarks}

To conclude, we considered a kinetic exchange opinion model in the presence of contrarian agents. These agents tend to take the opposite opinion of their social contacts. We considered a probability $a$ associated with the contrarian behavior, and with the complementary probability $1-a$ the interactions occurs through a kinetic exchange. For these interactions, we considered two distinct cases: pairwise and three-agent interaction.

Our target was to study the critical behavior of the model. For the case with pairwise interactions, we determined the stationary properties of the model through analytical calculations and numerical simulations. Our results indicate that the presence of contrarians leads to several consequences: (i) the absorbing state is destroyed, and in that place appears a paramagnetic phase, where the fractions of extreme opinions $+1$ and $-1$ are equal; (ii) the critical point is shifted when we increase $a$, decreasing the ordered phase; (iii) the three states ($+1$, $-1$ and $0$) coexist in the population in the ordered and disordered phases; (iv) the phase transition is suppressed for $a>a_{c}$. Our analytical prediction is $a_{c}\approx 0.068$, a value that was confirmed by the simulations. In other words, for $a>a_{c}$ there is no majority opinion in the society, independent of the other parameters of the model.

In the case of three-agent interactions, the impact of contrarians is more pronounced: (i) the absorbing phase is also destroyed; (ii) the transition is discontinuous only for $a<a_{TCP}$, presenting a tricritical point (TCP) at $a=a_{TCP}$; (iii) in this case, the transition becomes continuous for $a>a_{TCP}$; (iv) the transition is also suppressed for $a>a_{c}$. Our numerical results indicate that the TCP is located around $a_{TCP}\approx 0.034$, and the point above which there is no dominating opinion in the population was found to be $a_{c}\approx 0.068$, the same value observed in the model with pairwise interactions. 

For both cases, considering the transition at $a=a_{c}$, we verified numerically that the critical exponents are typical of the mean-field Ising model, suggesting a universal critical behavior drived by the contrarian agents, independent of the type of social interactions (pairwise or three agent interactions).


\section*{Acknowledgments}

The authors acknowledged financial support from the Brazilian scientific funding agencies CNPq and CAPES.

\end{document}